\documentclass[aps,prd,showpacs,preprintnumbers,nofootinbib,floatfix,floats,groupedaddress,twocolumn]{revtex4-1}

\usepackage{bm}
\usepackage{latexsym}
\usepackage{dcolumn}
\usepackage{amsmath,amsfonts,amssymb}
\usepackage{graphicx,epsfig}
\usepackage{color}
\usepackage[active]{srcltx}
\usepackage{hyperref}
\usepackage[caption=false]{subfig}

\DeclareSymbolFont{sfletters}{OML}{cmbrm}{m}{}
\DeclareMathSymbol{\smu}{\mathord}{sfletters}{"16}
\DeclareMathSymbol{\snu}{\mathord}{sfletters}{"17}
\usepackage{bm}

\usepackage{slashed}
\usepackage{mathrsfs}
\usepackage{tikz}
\usetikzlibrary{shapes,snakes}
\usepackage{dsfont}
\usepackage{mathtools}
\def\nn{\nonumber}

\def\l{\left}
\def\r{\right}
\def\DM{\mathrm{d}}

\newcommand{\gae}{\lower 3pt \hbox{$\,\, \buildrel {\scriptstyle >}\over {\scriptstyle
\sim}\,\,$}}
\newcommand{\lae}{\lower 2pt \hbox{$\, \buildrel {\scriptstyle <}\over {\scriptstyle
\sim}\,$}}

\reversemarginpar

\begin{document}

\title{Normal coordinates based on curved tangent space}

 \author{Hari K}
 \email{harik@physics.iitm.ac.in}
 \author{Dawood Kothawala}
 \email{dawood@iitm.ac.in}
 \affiliation{Department of Physics, Indian Institute of Technology Madras, Chennai 600 036}

\date{\today}
\begin{abstract}
\noindent
Riemann normal coordinates (RNC) at a regular event $p_0$ of a spacetime manifold $\mathcal{M}$ are constructed by imposing: (i) $g_{\textsf{ab}}|_{p_0}=\eta_{ab}$, and (ii) $\Gamma^\textsf{a}_{\phantom{\textsf a}\textsf{bc}}|_{p_0}=0$. There is, however, a third, {\it independent}, assumption in the definition of RNC which essentially fixes the {\it density of geodesics} emanating from $p_0$ to its value in flat spacetime, viz.: (iii) the tangent space $\mathcal{T}_{p_0}(\mathcal{M})$ is {\it flat}. 
We relax (iii) and obtain the normal coordinates, along with the metric $g_{\textsf{ab}}$, when 
$\mathcal{T}_{p_0}(\mathcal{M})$ is a maximally symmetric manifold $\widetilde{\mathcal M}_{\Lambda}$ with curvature length $|\Lambda|^{-1/2}$. 
In general, the ``rest" frame defined by these coordinates is non-inertial with an additional acceleration $\bm a = - ({\Lambda}/3) \, \bm x$ depending on the curvature of tangent space.

Our geometric set-up provides a convenient probe of local physics in a universe with a cosmological constant $\Lambda$, now embedded into the local structure of spacetime as a fundamental constant associated with a curved tangent space. We discuss classical and quantum implications of the same. 

\end{abstract}

\pacs{04.60.-m}
\maketitle
\vskip 0.5 in
\noindent
\maketitle
\section{Introduction} \label{sec:intro} 
\begin{figure}[!h]%
    {{\includegraphics[width=0.4\textwidth]{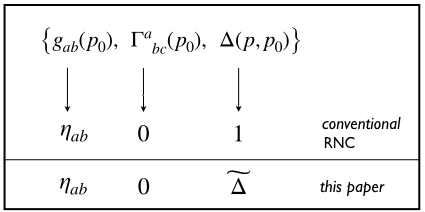} }}%
    \caption{Basic idea of this work; see also {\bf Fig. \ref{fig:exp-map}}.}%
    \label{fig:summary}%
\end{figure}

\begin{figure}[!htb]%
    {{\includegraphics[width=0.5\textwidth]{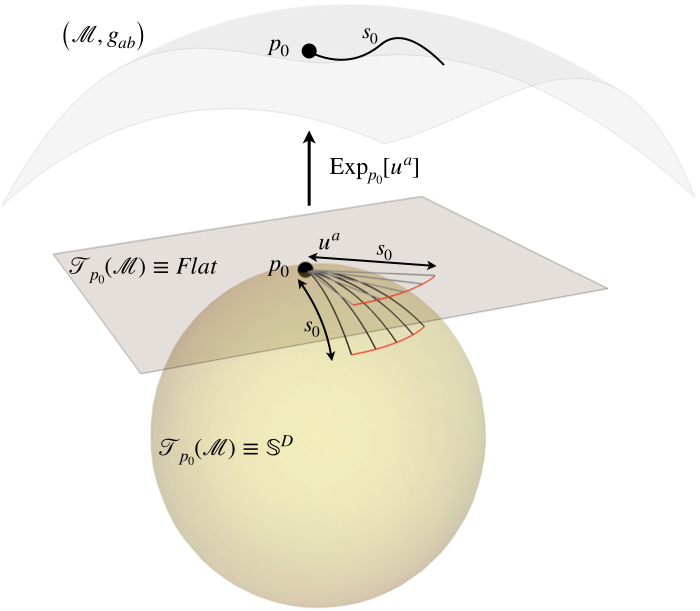} }}%
    \caption{Exponential map from $\mathcal{T}_{p_0}(\mathcal{M})$ to an open subset $\mathcal{U}$ of $\mathcal{M}$. If $\mathcal{T}_{p_0}(\mathcal{M})$ is 
    taken as an arbitrary maximally symmetric space $\widetilde{\mathcal M}_\Lambda$, one must properly account for the {\it density of geodesics} appropriate to $
    \widetilde{\mathcal M}_\Lambda$.}%
    \label{fig:exp-map}%
\end{figure}

Given a spacetime manifold with a metric, $\l( \mathcal{M}, \bm g \r)$, the most primitive structure that carries information about spacetime curvature at a (regular) event $p_0$ is perhaps the congruence of geodesics emanating from $p_0$. One can characterise the spacetime geometry in a convex normal neighbourhood of $p_0$ by assigning to points in this neighbourhood coordinates based on this congruence. These coordinates are known as Riemann normal coordinates (RNC) \cite{book-MTW,lb-riemann}, and, besides being a convenient computational tool, they provide a concrete realisation of the {\it principle of equivalence}. This last fact follows from the conditions characterizing the RNC : (i) $g_{ab}(p_0)=\eta_{ab}$, and (ii) $\Gamma^a_{\phantom{a}bc}(p_0)=0$. However, implicit in these conditions is the assumption that the tangent space $\mathcal{T}_{p_0}(\mathcal{M})$ one is using for the (inverse) exponential map (which eventually defines the RNC), is itself {\it flat}. In fact, the frame $\{ \bm e_{\textsf a}(p_0) \}$ that defines the RNC has its tetrad vectors normalised as $\bm{e}_{\textsf a} \bm \cdot \bm{e}_{\textsf b} = \eta_{\textsf {ab}}$. While this might seem like a {\it sufficient} condition to impose local flatness, it does not uniquely capture all possible information about the background spacetime. The reason for this is simple to see: RNC by definition are so constructed as to yield geodesics emanating from $p_0$ by ``straight lines". This, as we argue in this paper, is tied to the choice of a flat tangent space $\mathcal{T}_{p_0}(\mathcal{M})$. While a {\it single} geodesic connecting a point $p_0$ to $p$ (lying in the normal neighbourhood of $p_0$) suffices to assign to $p$ its RNC, a bunch of geodesics emanating from $p_0$ - {\it geodesic spray} - carries more information that is lost if the geodesics are modelled as straight lines as in RNC. We will make these statements more concrete in the rest of the paper, but for now, let us emphasise that this much at least is true: While any coordinates imposing conditions (i) and (ii) will ensure local flatness, one can still impose additional conditions on our choice of local coordinates such that the {\it density of geodesics} at $p_0$ is fixed not to its Minkowski value, but to a value set by an arbitrary maximally symmetric manifold $\widetilde{\mathcal M}_{\Lambda}$ with curvature length $|\Lambda|^{-1/2}$. This makes all the more sense in the backdrop of cosmology, since local physics in a universe with a non-zero cosmological constant $\Lambda$ would be described better by geodesic sprays modelled on, say, de Sitter rather that Minkowski spacetime. {\bf Figs. \ref{fig:summary}} and {\bf \ref{fig:exp-map}} give the basic idea.

            \begin{figure*}[!htb]%
                \centering
                \subfloat[The Synge world function $\Omega$ determines tangent vectors to geodesics connecting two points.]
                {{\includegraphics[width=0.45\textwidth]{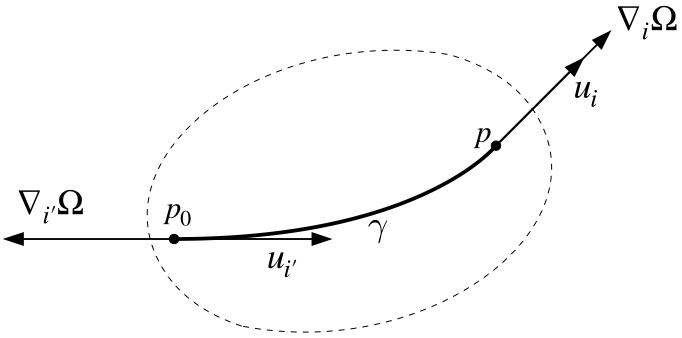} }}
		\subfloat[The van Vleck determinant $\Delta$ determines spread of geodesics.]
                {{\includegraphics[width=0.6\textwidth]{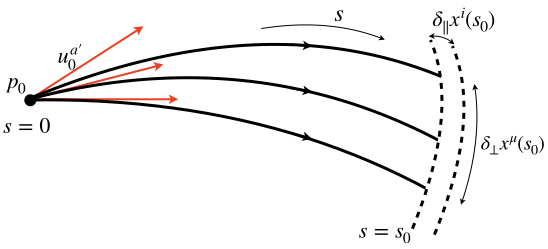} }}
                \caption{Geodesic spray from a given point $p_0$ is obtained from solution of the standard variational principle for geodesics by fixing the initial point as $p_0$ and varying the tangent vectors $u^a(0)$ at $p_0$. The {\it transverse} variation $\delta_{_\perp} x^\mu(s_0)$ provides a measure of (de-)focussing of the geodesics, and is characterised by the van Vleck determinant.}%
            \label{fig:geodesic-spray}%
            \end{figure*}

\textbf{{To summarise}:} The tangent space provides us with three basic geometrical objects, 
$\{ g_{\textsf{ab}}(p_0), \Gamma^{\textsf a}_{\phantom{\textsf{a}}\textsf{bc}}(p_0), {\widetilde{\Delta}}(p,p_0) \}$. Conventional RNC fixes ${\widetilde{\Delta}} = 1$, which, as we show here, is mathematically untenable. Since the covariant Taylor expansion of ${\widetilde{\Delta}}$ near $p=p_0$ starts at quadratic order (see text), our modification does not alter the metric and connection {\it at} $p_0$, but corrects it at $O(\textsf{x}^2)$.

\vspace{0.85cm}
\underline{\it Organisation of the paper}:
	In {\bf Sec.\,\ref{sec:geodesic-spray}}, we describe some geometrical aspects of geodesic sprays in arbitrary space(times), highlight the role of the van Vleck determinant, and bring into focus its role in defining local coordinates based on a maximally symmetric tangent space. In {\bf Sec.\,\ref{sec:arb-tan-sp}}, using the results from Sec.\,\ref{sec:geodesic-spray}, we introduce new locally inertial coordinates, and construct the corresponding metric tensor $g_{\textsf{ab}}$ to fourth order in series expansion. {\bf Sec.\, \ref{sec:max-symm-sp}} then applies the above formalism to a spacetime which is itself maximally symmetric. When the curvature scalar of this spacetime and the tangent space are equal, we show that the series for $g_{\textsf{ab}}$ can be {\it summed exactly}, and the resultant metric is precisely that of a maximally symmetric spacetime expressed in {\it embedding coordinates}!\cite{book-weinberg} We then proceed to discuss, in {\bf Sec.\;\ref{sec:applications}}, implications of our result for observables in classical and quantum physics, as well as in describing local physics in a universe wherein the cosmological constant $\Lambda$ is treated as a fundamental constant.

\underline{\it Notation}: 
We work in $D$ dimensions, and use the shorthand $D_k=D-k$. Latin and Greek indices denote spacetime and space components respectively. Latin indices in sans-serif font $\textsf{a, b \ldots}$ denote frame components, and will also be conveniently used to represent an object as an element of the tangent space. Further, since we will encounter many quantities that are bi-tensors depending on two points $p_0$ and $p$, we will often identify tensor indices at $p_0$ by a prime. Therefore, $Q^{abj'}$ denotes an object which is a vector at $p_0$ and a $(2,0)$ tensor at $p$, while $u^{\textsf{a}'}$ are frame components of a vector $u^{i'}$ at $p_0$, and also denotes a vector in $\mathcal{T}_{p_0}(\mathcal{M})$.

\section{Geodesic sprays, van Vleck determinant, and RNC} \label{sec:geodesic-spray} 
%
\subsection{Density of geodesics and the van Vleck determinant} \label{sec:density}
Our main focus in this section will be to review geometric aspects of geodesic spray emanating from an arbitrary (but regular) spacetime event $p_0$, in such a manner as to elucidate the connection with RNC. This later connection is what we will exploit in the subsequent sections to improvise on the conventional construction of RNC, an improvisation that, as we shall see, is best interpreted in terms of the geometry of the tangent space $\mathcal{T}_{p_0}(\mathcal{M})$ being maximally symmetric.

Let $\ell(x^{i'}, x^i)$ denote the length of the geodesic between any two points $p_0$, $p$ with coordinates $x^{i'}$ and $x^{i}$, in an arbitrary manifold. From standard variational analysis, we know that $\partial \ell / \partial x^{i'} = - \varepsilon u_{i'}$ and $\partial \ell / \partial x^{i} = + \varepsilon u_{i}$ (where $\varepsilon = u^2 = \pm 1$). See {\bf Fig.\,\ref{fig:geodesic-spray}} for the geometric setup and notations. Fixing $x^{i'}$ and varying $x^{i}$, we obtain $\delta \ell = \delta x^{i} \l( \partial \ell / \partial x^{i} \r) = \varepsilon u_{i} \delta x^{i}$. We ask if there is a one-to-one correspondence between $\delta x^{i}$ and $\delta u^{i'}$, so that one can trade off the variables $(x^{i'}, x^{i})$ with $(x^{i'}, u^{i'})$. It is easy to see that such a correspondence can indeed be established provided we restrict to variations $\delta x^i$ that are orthogonal to $u^i$, since the component of $\delta x^i$ along $u^i$ would simply shift the end-point along the same geodesic, and hence all such variations will correspond to the same $u^{i'}$. A non-trivial map between $\delta x^i$ and $\delta u^{i'}$ therefore exists only for variations orthogonal to $u^i$, which we denote by $\delta_\perp x^\mu$, and hence, for all these variations (from above), $\delta \ell=0$. Moreover, assuming the tangent vectors are all normalised to $\pm 1$, we have $u^{i'} \delta u_{i'}=0$, and hence the variations $\delta u_{i'}$ are orthogonal to $u^{i'}$.

To summarise, {\it for variations that keep constant the geodesic distance} of a point $p$ from a fixed point $p_0$, say $\ell(p_0, p)=s_0$, there is a one-to-one map between $\delta u_{i'}$ and $\delta x^i$. The Jacobian matrix corresponding to this map is given by \cite{mcg-chaos_in}
\begin{eqnarray}
	{\mathcal C}_{a' b}=\frac{\partial u_{a'}}{\partial x^{b}} 
= - \varepsilon \frac{\partial^2 \ell}{\partial  x^{a'} \partial x^{b}}
\end{eqnarray}
We call the set of events $p$ generated by such variations an {\it equi-geodesic surface}; some geometrical aspects of these surfaces are discussed in Appendix \ref{app:equi-geodesic}. While the matrix ${\mathcal C}_{\mu'\nu}$ comes with a natural interpretation, for dimensional reasons, it turns to be more convenient and insightful to define instead the matrix
\begin{eqnarray}
	{\mathcal D}_{a'b}=-\frac{1}{2} \varepsilon \frac{\partial^2 \ell^2}{\partial  x^{a'} \partial x^{b}}
\end{eqnarray}
whose determinant is closely related to the so called van Vleck determinant\cite{mv-van_vleck} (see below). The determinants of the matrices $\bm {\mathcal C}$ and $\bm {\mathcal D}$ are related in a simple manner. First, notice that 
\begin{eqnarray}
	{\mathcal D}_{a'b} = s_0 {\mathcal C}_{a'b} + \varepsilon u_{a'} u_{b}
\end{eqnarray}
Now, from discussion above, we know that $u^{a'}, u^{b}$ lie in the kernel of ${\mathcal C}_{ab'}$ (this can also be established by a quick computation), and 
${\mathcal D}_{a'b} u^{a'} = u_b, \; {\mathcal D}_{a'b} u^{b} = u_{a'}$. A straightforward computation then yields
\begin{eqnarray}
	\Delta(p_0, p) \coloneqq \frac{{\rm det} [{\mathcal D}_{a'b}]}{ \sqrt{|g(p_0)|} \sqrt{|g(p)|} } = \frac{s_0^{D-1} {\rm det} [{\mathcal C}_{\mu' \nu}]}{ \sqrt{|h(p_0)|} \sqrt{|h(p)|} }
\label{eq:vvd-jacobian-txt}
\end{eqnarray}
which is the desired relation between the Jacobian of transformation from final end-point to the initial tangent vector and the so called van Vleck bi-scalar $\Delta(p_0, p)$ which is defined by the LHS above. A short derivation of this is sketched in the Appendix \ref{app1}. (Note that the derivation naturally yields correct factors of metric determinants so as to give a relation between scalars.)
\subsection{Normal coordinates based on the Jacobian matrix ${\mathcal C}_{ab'}$}\label{sec:normal}
Riemann normal coordinates of a point $p$ in a convex normal neighbourhood of $p_0$ are given by \cite{ep-ap-iv-the_motion}
\begin{eqnarray}
	x^{\textsf a} = - \eta^{\textsf{ab}} e^{i'}_{\textsf b} \nabla_{i'} \Omega
\end{eqnarray}
where $e^{i'}_{\textsf b}$ is an orthonormal tetrad at $p_0$: $\bm e_{\textsf b} \cdot \bm e_{\textsf b} = \eta_{\textsf ab}$, and $\Omega(p_0, p) = \sigma^2(p_0, p)/2$ is the Synge world function \cite{book-jls}. In terms of frame components of the tangent vector, this is equivalent to assigning to $p$ the coordinates
\begin{eqnarray}
	x^{\textsf a} = s \eta^{\textsf{ab}} u_{\textsf b'}
\end{eqnarray}
where $s = \sqrt{|\sigma^2|}$ is the geodesic length between $p_0$ and $p$. 

To see the connection with the discussion in earlier subsection, vary $u_{\textsf b'}$, and track the corresponding variations of $x^{\textsf a}$ keeping $s$ constant: $s=s_0$. This gives 
\begin{eqnarray}
	\delta_\perp x^{\mu} = s_0 \eta^{\mu \nu} \delta u_{\nu'}
\label{eq:delx-rnc}
\end{eqnarray}
Therefore, in RNC, the volume spanned by the variations $\delta x^{\mu}$ is given by ${\rm det} \l[ s_0 \delta^{\mu \nu} \r] = s_0^{D-1}$.
On the other hand, from the previous subsection, we have
\begin{eqnarray}
	{\mathcal C}_{a' b}=\frac{\partial u_{a'}}{\partial x^{b}} 
\implies
\delta x^{\mu} = \l[C^{-1}\r]^{\nu' \mu} \delta u_{\nu'}
\end{eqnarray}
so that the volume spanned by the variations $\delta x^{\mu}$ is given by ${\rm det} [{\mathcal C}^{-1 \; \mu \nu'}]$, appropriately scalarised. From the previous subsection -- see Eqs.~(\ref{eq:vvd-jacobian-txt}) -- this is given by $s_0^{D-1} \widetilde{\Delta}^{-1}$, and hence, given a fixed set of initial variations $\delta u_{\nu'}$ at $p_0$, the above two volumes are clearly different in general, unless $\widetilde{\Delta}=1$, that is, for flat spacetime. This is a direct consequence of the fact that the tangent space $\mathcal{T}_{p_0}(\mathcal{M})$ is assumed flat, since the assignment $x^{\textsf a} = s \eta^{\textsf{ab}} u_{\textsf b'}$, along with the identity $\eta_{\textsf{ab}} x^{\textsf a} x^{\textsf b} = \sigma^2$, ensures that geodesics based on these coordinates will be straight lines. For a non-flat tangent space, previous discussion immediately yields the following corrected definition of coordinates
\begin{eqnarray}
	\hat{x}^{\textsf a} &=& \Biggl\{ {\rm det} [{\mathcal C}^{-1}_{\mu \nu'}] \Biggl\}^{1/(D-1)} \eta^{\textsf{ab}} u_{\textsf b'}
\nn \\
	&=& s_0 \widetilde{\Delta}^{-1/D_1} \eta^{\textsf{ab}} u_{\textsf b'}
\nn \\
	&=& - \eta^{\textsf{ab}} e^{i'}_{\textsf b} (\nabla_{i'} \Omega) \; \widetilde{\Delta}^{-1/D_1} \nn \\
	&=& -e^{\textsf{a}}_{i'}\Omega^{i'}\widetilde{\Delta}^{-1/D_1}
\label{eq:def-coord}
\end{eqnarray}
that ensures that the volume spanned by the variations $\delta x^{\mu}$ now correctly takes into account the Jacobian ${\rm det} [{\mathcal C}^{-1 \; \mu \nu'}]$. As for the variations $\delta_\parallel x^i$, these satisfy the geodesic equation, as they should. This is easily verified by an explicit computation, using the Christoffel symbols for the metric which we will derive in the next section. (A formal proof of this is straightforward.)

We now have the appropriate generalisation of Riemann normal coordinates to the case when the tangent space is non-flat. The curvature of the tangent space is captured by its van Vleck determinant $\widetilde{\Delta}$, and a flat tangent space (which is the standard case) has $\widetilde{\Delta}=1$, reproducing the conventional RNC. Now, given a fixed set of initial variations $\delta u_{\nu'}$ at $p_0$ equal to the number of degrees of freedom $N_{\rm dof}$ of the system, ${\rm det} [{\mathcal C}_{\mu \nu'}]$ acquires the interpretation of {\it density of geodesics} emanating from $p_0$. Therefore, we have essentially obtained the normal coordinates that carry the correct information about {\it density of geodesics} appropriate to a non-flat tangent space. We next proceed to derive the metric in these coordinates.

\section{Metric in the generalised RNC} \label{sec:arb-tan-sp}

Once an orthornormal tetrad is fixed at $p_0$, the differential of coordinates of a point $p$ in the neighbourhood of $p_0$ are related by
\begin{eqnarray}
	\DM\hat{x}^{\textsf{a}} &=\left[-\widetilde{\Delta}^{-\frac{1}{D_{1}}}e^{\textsf{a}}_{i'}\Omega^{i'}_{j}-\frac{\hat{x}^{\textsf{a}}}{D_{1}}\partial_{j}\ln\left(\widetilde{\Delta}\right)\right]\DM x^{j}
\label{coord-var}
\end{eqnarray}
The coincidence limit is easily shown to give
\begin{eqnarray}
	\left[\frac{\partial\hat{x}^{\textsf{a}}}{\partial {x}^{i}}\right]_{p_0} =e^{\textsf{a}}_{i'}(p_0)  \hspace{0.5cm}; \hspace{0.5cm} \left[\frac{\partial \hat{x}^{i}}{\partial {x}^{\textsf{a}}}\right]_{p_0} =e^{i'}_{\textsf{a}}(p_0)
\label{coord-trans}
\end{eqnarray}
\\
The line element can now be expressed as $\DM\sigma^2=g_{ij}\DM x^{i}\DM x^{j}=\eta_{\textsf{a}\textsf{b}}e^{\textsf{a}}_{i}e^{\textsf{b}}_{j}\DM x^{i} \DM x^{j}=g_{\textsf{a}\textsf{b}}\DM \hat{x}^{\textsf{a}} \DM \hat{x}^{\textsf{b}}$, where $g_{ij}$ is the metric of the background spacetime in arbitrary coordinates and $g_{\textsf{ab}}$ is the metric in normal coordinates, which we now evaluate. As we will show, $g_{\textsf{ab}}$ depends on the curvature of the background spacetime as well as that of the tangent space.
\\

Using the series expansions given in Ref. \cite{smc-vacc}, Eq.~(\ref{coord-var}) can be inverted to obtain $e^{\textsf{a}}_{i}$, and the line element can be evaluated in terms of $x^{\textsf a}$, from which the metric can finally be read off. The derivation involves some subtlety since $\widetilde{\Delta}$ also appears explicitly in the definition of coordinates; see Eq.~(\ref{eq:def-coord}). Relevant details are given in Appendix \ref{app2}. 

The final metric, which is our key result, is given by:
\begin{widetext}
	\begin{align}
	\label{eq:metric-full}
		g_{\textsf{a}\textsf{b}} &= \eta_{\textsf{a}\textsf{b}}
		+ \frac{1}{3}\left( -R_{\textsf{a}\textsf{c}\textsf{b}\textsf{d}}+\frac{1}{D_{1}}\eta_{\textsf{a}\textsf{b}}\widetilde{R}_{\textsf{c}\textsf{d}}+\frac{2}{D_{1}}\eta_{\textsf{c}\textsf{(a}}\widetilde{R}_{\textsf{b)}\textsf{d}} \right)\widehat{x}^{\textsf{c}}\widehat{x}^{\textsf{d}}
	+ Q_{\textsf{abcde}} \widehat{x}^{\textsf{c}} \widehat{x}^{\textsf{d}} \widehat{x}^{\textsf{e}}
	+ Q_{\textsf{abcdef}} \widehat{x}^{\textsf{c}} \widehat{x}^{\textsf{d}} \widehat{x}^{\textsf{e}} \widehat{x}^{\textsf{f}}
        + O(\textsf{x}^5)
         \\
        \nn \\
         Q_{\textsf{abcde}} &= -\frac{1}{6}R_{\textsf{a}\textsf{c}\textsf{b}\textsf{d};\textsf{e}} -\frac{1}{6D_{1}}\eta_{\textsf{a}\textsf{b}}\widetilde{R}_{\textsf{c}\textsf{d};\textsf{e}}+ \frac{1}{2D_{1}}\widetilde{R}_{\textsf{c}\textsf{d};(\textsf{a}}\eta_{\textsf{b})\textsf{e}}
         \nn \\
         Q_{\textsf{abcdef}} &= 
         \underbrace{
         -\frac{1}{20}R_{\textsf{a}\textsf{c}\textsf{b}\textsf{d};\textsf{e}\textsf{f}}+\frac{2}{45}R_{\textsf{a}\textsf{c}\textsf{m}\textsf{d}}R^{\textsf{m}}_{\phantom{\textsf{m}}\textsf{e}\textsf{b}\textsf{f}}
         }_{\rm{standard~RNC}}
         ~~-~~
         \underbrace{
         \frac{2}{9D_{1}}R_{\textsf{a}\textsf{c}\textsf{b}\textsf{d}}\widetilde{R}_{\textsf{e}\textsf{f}}-\frac{2}{9D_{1}}\widetilde{R}_{\textsf{m}\textsf{c}} R^{\textsf m}_{\phantom{\textsf{m}}\textsf{d}\textsf{e}(\textsf{a}}\eta_{\textsf{b})\textsf{f}}
         }_{\rm{coupled~ terms}}         
         ~~ + \frac{1}{20D_{1}}\eta_{\textsf{a}\textsf{b}}\widetilde{R}_{\textsf{c}\textsf{d};\textsf{e}\textsf{f}} +\frac{1}{5D_{1}}\widetilde{R}_{\textsf{c}\textsf{d};\textsf{e}(\textsf{a}}\eta_{\textsf{b})\textsf{f}}
         \nn \\ 
		& \quad
		%
		+\frac{1}{6D_{1}^2}\eta_{\textsf{a}\textsf{b}}\widetilde{R}_{\textsf{c}\textsf{d}}\widetilde{R}_{\textsf{e}\textsf{f}}+\frac{2}{3D_{1}^2}\eta_{\textsf{c}\textsf{(a}}\widetilde{R}_{\textsf{b)}\textsf{d}}\widetilde{R}_{\textsf{e}\textsf{f}}
		+\frac{1}{9D_{1}^2}\eta_{\textsf{e}\textsf{f}}\widetilde{R}_{\textsf{a}\textsf{c}}\widetilde{R}_{\textsf{b}\textsf{d}} +\frac{1}{90D_{1}}\eta_{\textsf{a}\textsf{b}}\widetilde{R}^{\textsf{k}}_{\phantom{\textsf{m}}\textsf{c}\textsf{m}\textsf{d}}\widetilde{R}^{\textsf{m}}_{\phantom{\textsf{m}}\textsf{e}\textsf{k}\textsf{f}}+\frac{2}{45D_{1}}\eta_{\textsf{c}\textsf{(a}}\widetilde{R}^{\textsf{k}}_{\phantom{\textsf{k}}\textsf{b)}\textsf{m}\textsf{d}}\widetilde{R}^{\textsf{m}}_{\phantom{\textsf{m}}\textsf{e}\textsf{k}\textsf{f}} 
\nn 
	\end{align}
\end{widetext}
where $R_{\textsf{a}\textsf{c}\textsf{b}\textsf{d}}$ and $\widetilde{R}_{\textsf{a}\textsf{c}\textsf{b}\textsf{d}}$ are the Riemann tensors of the background space and the tangent space respectively and the brackets in lower indices indicate that they are symmetrised. The indices of these tensors are lowered and raised using $\eta_{\textsf{a}\textsf{b}}$ and $\eta^{\textsf{a}\textsf{b}}$ since they are evaluated at the base point.

\section{Fixing ``density of geodesics" using maximally symmetric tangent space} \label{sec:max-symm-sp}

Throughout this paper, we will quote all the expressions keeping the curvature tensors $\widetilde{R}_{\textsf{a}\textsf{c}\textsf{b}\textsf{d}}$ associated with the tangent space arbitrary; in particular, one could simply set $\widetilde{R}_{\textsf{a}\textsf{c}\textsf{b}\textsf{d}} \to R_{\textsf{a}\textsf{c}\textsf{b}\textsf{d}}$. This is equivalent to
setting $\widetilde{\Delta}=\Delta$, which is a perfectly acceptable (and even more general) choice, and makes no reference to a curved tangent space at all. We comment further on this point of view in {\bf Sec.\;\ref{sec:implications}}, $2^{\rm nd}$ paragraph. 

However, keeping with the spirit of conventional RNC, our key interest is in {\it modelling} the tangent space by a maximally symmetric manifold $\widetilde{\mathcal M}_{\Lambda}$ of dimension $D$, $|\widetilde{\Lambda}|^{-1/2}$ being the curvature length scale determined by the parameter $\widetilde{\Lambda}$. In fact, since there generically is no canonical way to identify points between two different manifolds, it would not make much sense to choose an arbitrary manifold to model the tangent space, since the connection between tangent spaces at two different points is then unclear. For maximally symmetric case, the connection is simply an element of the symmetry group of the manifold. 

For the maximally symmetric tangent space, the metric in Eq.~(\ref{eq:metric-full}) can be reduced by using the definitions,
\begin{eqnarray}
	\widetilde{R}_{abcd} &=&\widetilde{\Lambda} \left(\widetilde{g}_{ac}\widetilde{g}_{bd}-\widetilde{g}_{ad}\widetilde{g}_{bc}\right) \nn \\
	\widetilde{R}_{ab} &=& D_{1} \widetilde{\Lambda} \widetilde{g}_{ab}\\
	\widetilde{R} &=& DD_{1} \widetilde{\Lambda} \nn
\end{eqnarray}
where $\widetilde{g}_{ab}$ is the metric of maximally symmetric space which we assume that reduces to $\eta_{ab}$ at the base point. Using these in Eq.~(\ref{eq:metric-full}), we obtain
\begin{widetext}
\begin{equation}
	\begin{aligned}
		g_{\textsf{a}\textsf{b}} &=\eta_{\textsf{a}\textsf{b}}+\frac{1}{3}\left( -R_{\textsf{a}\textsf{c}\textsf{b}\textsf{d}}+\widetilde{\Lambda}\eta_{\textsf{a}\textsf{b}}\eta_{\textsf{c}\textsf{d}}+2\widetilde{\Lambda}\eta_{\textsf{a}\textsf{c}}\eta_{\textsf{b}\textsf{d}} \right)\widehat{x}^{\textsf{c}}\widehat{x}^{\textsf{d}}-\frac{1}{6}R_{\textsf{a}\textsf{c}\textsf{b}\textsf{d};\textsf{e}}\widehat{x}^{\textsf{c}}\widehat{x}^{\textsf{d}}\widehat{x}^{\textsf{e}} 
		\\
		& \hspace{0.9cm} + \left( -\frac{1}{20}R_{\textsf{a}\textsf{c}\textsf{b}\textsf{d};\textsf{e}\textsf{f}}+\frac{2}{45}R_{\textsf{a}\textsf{c}\textsf{l}\textsf{d}}R^{\textsf{l}}_{\phantom{\textsf{l}}\textsf{e}\textsf{b}
		\textsf{f}} 
		-\frac{2\widetilde{\Lambda}}{9}\eta_{\textsf{e}\textsf{f}}R_{\textsf{a}\textsf{c}\textsf{b}\textsf{d}}+\frac{8\widetilde{\Lambda}^2}{45}\eta_{\textsf{a}\textsf{b}}\eta_{\textbf{c}\textsf{d}}\eta_{\textsf{e}\textsf{f}}+\frac{37\widetilde{\Lambda}^2}{45}\eta_{\textsf{a}\textsf{c}}\eta_{\textsf{b}\textsf{d}}\eta_{\textsf{e}\textsf{f}} \right)\widehat{x}^{\textsf{c}}\widehat{x}^{\textsf{d}}\widehat{x}^{\textsf{e}}\widehat{x}^{\textsf{f}}+O({\textsf x}^5)\;
	\end{aligned}
\label{metric-ms}
\end{equation}
\end{widetext}
We can obtain the above metric via alternate route, by using the geodesic equation and its higher derivatives along with the known expression for van Vleck determinant of maximally symmetric spacetimes
\begin{eqnarray}
\widetilde \Delta^{-\frac{1}{D_{1}}}=\l(\frac{\sin{\l(s\sqrt{|\widetilde{\Lambda}|}\r)}}{s\sqrt{|\widetilde{\Lambda}|}},0,\frac{\sinh{\l(s\sqrt{|\widetilde{\Lambda}|}\r)}}{s\sqrt{|\widetilde{\Lambda}|}}\r)
\label{eq:vvd-max-symm}
\end{eqnarray} 
of positive, zero and negative curvature, respectively \cite{equi-geod}. 

A particularly interesting case is when the the background spacetime is itself maximally symmetric with its own constant $\Lambda$, in which case the above metric becomes
\begin{equation}
	\begin{aligned}
		g_{\textsf{a}\textsf{b}} &=\eta_{\textsf{a}\textsf{b}}+\frac{1}{3}\left(\left[ \widetilde{\Lambda}-\Lambda \right]\eta_{\textsf{a}\textsf{b}}\eta_{\textsf{c}\textsf{d}}+\left[ 2\widetilde{\Lambda}+\Lambda \right]\eta_{\textsf{a}\textsf{c}}\eta_{\textsf{b}\textsf{d}}\right)\widehat{x}^{\textsf{c}}\widehat{x}^{\textsf{d}}\\
		& \quad +\frac{1}{45}\left( \left[ 2\Lambda^2+8\widetilde{\Lambda}^{2}-10\widetilde{\Lambda}\Lambda \right]\eta_{\textsf{a}\textsf{b}}\eta_{\textsf{c}\textsf{d}}\eta_{\textsf{e}\textsf{f}} \right. \\
	& \quad \; \left. +\left[ -2{\Lambda}^{2}+37\widetilde{\Lambda}^{2}+10\widetilde{\Lambda}\Lambda \right]\eta_{\textsf{a}\textsf{c}}\eta_{\textsf{b}\textsf{d}}\eta_{\textsf{e}\textsf{f}} \right)\widehat{x}^{\textsf{c}}\widehat{x}^{\textsf{d}}\widehat{x}^{\textsf{e}}\widehat{x}^{\textsf{f}}\\
		& \quad +O(\textsf{x}^{5}) \; 
	\end{aligned}
\label{metric-ms-ms1}
\end{equation} 
We now point out a remarkable feature of the above metric, which we justify {\it a posteriori}. Define $\xi \equiv ({\widetilde{\Lambda}-\Lambda})/{2}$ and 
$\Lambda_{\rm{eff}} \equiv ({\widetilde{\Lambda}+\Lambda})/{2}$. Then, a set of terms in the above metric expansion can be {\it summed exactly}, and we obtain
\begin{equation}
	\begin{aligned}
		g_{\textsf{a}\textsf{b}} &=\eta_{\textsf{a}\textsf{b}} + 
		\frac{\Lambda_{\rm{eff}} }{1-\Lambda_{\rm{eff}} \eta_{\textsf{e}\textsf{f}}\widehat{x}^{\textsf{e}}\widehat{x}^{\textsf{f}}} \eta_{\textsf{a}\textsf{c}}\eta_{\textsf{b}			\textsf{d}}\widehat{x}^{\textsf{c}}\widehat{x}^{\textsf{d}} 
		+ F\left( \xi, \Lambda_{\rm{eff}} \right) \; 
	\end{aligned}
\label{metric-ms-ms}
\end{equation}
where $F\left( \xi, \Lambda_{\rm{eff}} \right)$ is a function that satisfies $F\left( 0, \Lambda_{\rm{eff}} \right)=0$, but can not otherwise be obtained in a closed form. 

The justification of this easily follows from the following observation: For maximally symmetric spacetimes (with curvature constant $\mathcal K$) expressed in embedding coordinates, the solutions of geodesic equation (with the starting point chosen as origin) are of the form $\sin\l( s \sqrt{|\mathcal K|} \r)$ or $\sinh\l( s \sqrt{|\mathcal K|}\r)$ \cite{book-weinberg}. Comparing this with the form of the van Vleck determinant, Eq.~(\ref{eq:vvd-max-symm}), we immediately see that for a maximally symmetric background spacetime with $\Lambda=\mathcal K=\widetilde{\Lambda}$, our coordinates reduce precisely to the embedding coordinates. The corresponding metric must then also reduce to the form given in Ch. 13 of  \cite{book-weinberg}, which is precisely given by the first two terms on RHS of (\ref{metric-ms-ms}). One must therefore have $F(0, \Lambda_{\rm{eff}})=0$.

The above observation provides a curious interpretation for the new coordinates that we have defined. As just shown, for maximally symmetric backgrounds, our coordinates $\widehat{x}^{\textsf{a}}$ can be interpreted as the embedding coordinates, with the embedding space a $(D+1)$ dimensional flat space(time). We have therefore essentially generalised the conventional $D$ dimensional RNC, based on a flat tangent space, by another set of ``flat" coordinates which are now inherited from a $(D+1)$ dimensional flat space(time). It is then no surprise that our new coordinates incorporate the correct density of geodesics, since the embedding map is smooth. Of course, the above interpretation in terms of embedding coordinates is very specific to background spacetimes which are themselves maximally symmetric, since embedding of an arbitrary manifold in a flat spacetime (of higher dimensions) will generically be more complicated and not amenable to any such nice interpretation.

{\bf Aside:} As a bonus, the above observation provides a slick way to obtain an exact expression for the metric of maximally symmetric spacetimes in Riemann normal coordinates, which can be derived by more conventional methods; see, for instance, Ref \cite{RNC_summed_for_ms}. We hope to elaborate on this elsewhere.
%
\section{Applications} \label{sec:applications} 
In this section, we sketch a few immediate implications and applications of the local metric that we have derived, making appropriate comments in the respective sub-sections below. We will restrict our discussion in this section to leading order terms in curvature, and hence ignore $O(\textsf x^2)$ term. Needless to say, many more applications can be discussed, and we hope the ones we discuss below will provide a motivation for future work along these lines.
%
\subsection{Acceleration of ``rest" observers} \label{sec:app-acceleration}
The best way to understand the significance of any coordinate chart $(t, x^{\mu})$ covering a region of spacetime is to study observers that are at rest with respect to the chart; that is, the $x^{\mu}=$ constant = $l^{\mu}$ observers. The {\it frame of reference} of such observers (that is, their rest frame) will then be inertial if their acceleration vanishes. In RNC coordinates, such ``static" observers are easily shown to have an acceleration
\begin{equation}
	a^0=0 \hspace{0.5cm}; \hspace{0.5cm} a^{\mu} = \frac{2}{3} R^{\mu}_{\phantom{\mu}0\nu0} l^{\nu}
\end{equation}
Thus, at the origin $x^\mu=0$, the frame is inertial, though in general, it is non-inertial. This is not difficult to understand since, given an orthonormal tetrad $e^i_{\textsf a}$ at $p_0$, one and only one tangent vector $\bm u(p_0)$ will coincide with $\bm e_{\textsf 0}$, and for the the point identified with this tangent vector, RNC will assign coordinates $l^{\mu}=0$.

We can do a similar analysis in our new coordinates. Consider, then, an observer on the trajectory $\widehat{z}^{\textsf i}(\tau)=(z^0(\tau), l^{\mu})$, 
with $l^{\mu}=$ constant. The four velocity associated with this observer is $\widehat{u}^{\textsf i}=\l[ (-g_{00})^{-1/2},0,0,0 \r]$, from which one can compute the acceleration 
$\widehat{a}^{\textsf k}\equiv \widehat{u}^{\textsf i} \nabla_{\textsf i} \widehat{u}^{\textsf k}$ to first order in curvature. The result gives
\begin{equation}
\widehat{a}^0=0 \hspace{0.5cm}; \hspace{0.5cm} \widehat{a}^\mu = \frac{2}{3} R^{\mu}_{\phantom{\mu}0\nu0} l^{\nu} + \frac{1}{3D_{1}} \widetilde{R}_{00} l^{\mu}
\end{equation}
\underline{\it Acceleration in FLRW universe}: It is of interest to explicit write down the above acceleration for the background spacetime describing an expanding universe. In the FLRW metric, we may use the canonical orthonormal basis at an arbitrary point $p_{0}$, and it is easily shown that $R^{\mu}_{\phantom{\mu}0\nu0}=-({\ddot{a}}/{a}) \delta^\mu_\nu$, where $a(t)$ is the scale factor and $\ddot{a}=\DM^2 a/\DM t^2$. Imposing Einstein equations with energy density $\rho$ and pressure $p$ gives:
\begin{eqnarray}
	\frac{\ddot{a}}{a}=-4\pi G \l( \frac{\rho}{D_{1}}+p \r)
\end{eqnarray}
We then see that the observers that are ``at rest" as dictated by our coordinates have acceleration
\begin{eqnarray}
	\widehat{a}^{\mu} &=& \l[\frac{8\pi G}{3}\l( \frac{\rho}{D_{1}}+p \r)-\frac{1}{3}\widetilde{\Lambda}\r]l^{\mu}
	\nn \\
	&=& \frac{8\pi G}{3}\l( \frac{\rho_{\rm m}}{D_{1}}+{p_{\rm m}} \r) l^{\mu}
		- \l[\frac{8\pi G}{3} \frac{D_2}{D_1} \rho_{\rm DE} + \frac{1}{3}\widetilde{\Lambda}\r]l^{\mu}
	\nn \\
\end{eqnarray}
where in the second equality we have separated the term with equation of state: $p_{\rm DE}=- \rho_{\rm DE}$ from the other sources $(\rho_{\rm m}, p_{\rm m})$. Thus, we see that our choice of coordinates based on a non-flat tangent space yields a natural set of ``rest" observers whose acceleration has a contribution from $\widetilde{\Lambda}$. Classically, none of this seems surprising, but quantum mechanically, the choice of coordinates does get tied with the choice of vacuum, and hence, the above result will have implications for vacuum energy and its interpretation as a cosmological constant. Needless to say, much more careful analysis would be needed to elaborate further on this. 

\subsection{The surface term in the Einstein-Hilbert action} \label{sec:app-Pc}
We have seen above that the rest frame identified by our coordinates has a contribution to its acceleration which, to the leading order, is directly proportional to the Ricci tensor of the model tangent space. It is then natural to look for similar effects on other observables of interest, particular the ones which depend on the choice of observers. The question one is interested in is the following: {\it Do all such objects and/or observables acquire correction from tangent space geometry}? 

To address this question, our next choice is to look at the structure of the Einstein-Hilbert (EH) lagrangian, since, as is well known, the only term in the EH lagrangian that can not be set to zero in a frame in which $\Gamma^{\textsf{a}}_{\phantom{\textsf{a}}\textsf{b}\textsf{c}}(p_0)=0$, is the surface term. More specifically, the EH action has the structure
$
R \sqrt{-g} = \l(\mathrm{bulk~part}\r) + \partial_{\textsf c} \l( \sqrt{-g} P^{\textsf c} \r)
$ \cite{book-tp}, where
\begin{eqnarray}
P^{\textsf{c}}  &=& {(-g)}^{-1} \partial_{\textsf{b}} \l[ (-g) g^{\textsf{b}\textsf{c}} \r]
 \nn \\
&=& \sqrt{-g} \left[ g^{\textsf{i} \textsf{k}} \Gamma^{\textsf{c}}_{\phantom{\textsf{c}}\textsf{i} \textsf{k}} - g^{\textsf{i} \textsf{k}} \Gamma^{\textsf{m}}_{\phantom{\textsf{m}}\textsf{k} \textsf{m}} \right] 
\label{eq:Pc}
 \end{eqnarray}
Although coordinate dependent, $P^{\textsf{c}}$ can be written in a covariant but observer dependent form. This is the reason why it is of considerable interest in the study of thermodynamics associated with local causal horizons, as its structure (for a foliation defined by carefully chosen observers) turns out to have information about the entropy associated with such horizons. We will now calculate this term in the local coordinates we have constructed. 
 
From the expression for the inverse metric and the Christoffel symbols given in Appendix \ref{app3}, it is straightforward to obtain
\begin{subequations}
	\begin{align}
		\begin{split}
			g^{\textsf b \textsf c}\Gamma^{\textsf a}_{\phantom{\textsf{a}}\textsf b \textsf c} &=\frac{2}{3}\eta^{\textsf a \textsf i}R_{\textsf i \textsf e} \widehat{x}^{\textsf e}+\frac{1}{3D_{1}}\left( \delta^{\textsf a}_{\textsf e}\eta^{\textsf b \textsf c}\widetilde{R}_{\textsf b \textsf c}+3\eta^{\textsf a \textsf i}\widetilde{R}_{\textsf i \textsf e} \right)\widehat{x}^{\textsf e} \\
	& \quad +O(\textsf{x}^2) \; 
		\end{split} \\
		\begin{split}
  			g^{\textsf a \textsf b}\Gamma^{\textsf c}_{\phantom{\textsf{c}}\textsf b \textsf c} &=-\frac{1}{3}\eta^{\textsf a \textsf b}R_{\textsf b \textsf e}\widehat{x}^{\textsf e}+\frac{1}{3D_{1}}\left(D+2\right)\eta^{\textsf a \textsf b}\tilde{R}_{\textsf b \textsf e}\widehat{x}^{\textsf e}+O(\textsf{x}^2)
		\end{split}
	\end{align}
\end{subequations}
where $\eta_{\textsf a \textsf b}$ is used to raise and lower the indices (this is valid at the leading order). Substituting into Eq.~(\ref{eq:Pc}), we obtain, to leading order,
\begin{eqnarray}
	P^{\textsf  a} &=& \Biggl[ \eta^{\textsf a \textsf k}R_{\textsf k \textsf m}+\frac{1}{3D_{1}}\left(\delta^{\textsf a}_{\textsf m}\eta^{\textsf i \textsf j}\tilde{R}_{\textsf i \textsf j}-D\eta^{\textsf a \textsf k}\widetilde{R}_{\textsf k \textsf m}\right) \Biggl] \hat{x}^{\textsf m}
\nn \\
\nn \\
\nn \\
	 &=& \Biggl[ R^{\textsf a}_{\phantom{\textsf a}\textsf m}-\frac{D}{3D_{1}} \left(\widetilde{R}^{\textsf a}_{\phantom{\textsf a}\textsf m}-\frac{1}{D} \delta^{\textsf a}_{\textsf m} \widetilde{R}\right) \Biggl] \hat{x}^{\textsf m}
\end{eqnarray}
Remarkably, the contribution from the tangent space comes as the traceless part of its Ricci tensor, and hence vanishes identically for maximally symmetric tangent spaces since for these, $\widetilde{R}^{\textsf a}_{\phantom{\textsf a}\textsf m} = (\widetilde{R}/D)  \delta^{\textsf a}_{\phantom{\textsf a}\textsf m}$. 

Here, then, is an instructive example of an object which does not depend on the tangent space geometry {\it as long as it is maximally symmetric}. It is unclear as of now whether $P^{\textsf c}$ vanishes to the next or higher orders as well. If it does, the mathematical reason behind it would be worth investigating in detail, given the role of $P^{\textsf c}$ in the thermodynamic aspects of gravity.

\section{Discussion and Implications} \label{sec:implications}
The main motivation of this work is simple: RNC, which serve as a very powerful local probe of spacetime geometry, are constructed assuming that, at some chosen point $p_0$: (i) $g_{ab}(p_0)=\eta_{ab}$, (ii) $\Gamma^a_{\phantom{a}bc}(p_0)=0$, and (iii) the geodesics emanating from $p_0$ are ``straight lines" of the Minkowski spacetime. This last condition essentially requires the coordinates to be characterised by the exponential map from $\mathcal{T}_{p_0}(\mathcal{M})$ to an open subset $\mathcal{U}$ of $\mathcal{M}$, with $\mathcal{T}_{p_0}(\mathcal{M})$ itself having the geometry of Minkowski spacetime. As we have highlighted in this work, this last condition (iii)  is unrelated to (i) and (ii). We can have (i) and (ii) while instead choosing the geodesics to be those corresponding to one of the homogenous geometries - maximally symmetric space(time)s - of which Minkowski spacetime is the simplest choice. In presence of the very strong evidence that we inhabit a universe with a positive cosmological constant $\Lambda>0$, it may not, however, be the best choice. While it is a mathematical theorem that a metric on {\it any} differentiable manifold can be expanded about the Minkowski spacetime, this does not imply, nor is it implied by, the fact that coordinates be defined based on a flat tangent space geometry. Since normal coordinates are based on geodesics from $p_0$, and curvature affects how geodesics deviate as they move away from $p_0$, geodesic normal coordinates based on a curved tangent space would differ from conventional RNC. As we have shown, the difference is characterised by the van Vleck determinant $\widetilde{\Delta}(p_0, p)$ of the tangent space. 

At this point, it is worth emphasising an important conceptual point. The $\widetilde{\Delta}(p_0, p)$ that appears in our definition is something that arises upon correctly incorporating the density of geodesics emanating from a point $p_0$. We fixed this density using as tangent space a maximally symmetric geometry $\widetilde{\mathcal M}_{\Lambda}$, and putting for $\widetilde{\Delta}$ the expression corresponding to $\widetilde{\mathcal M}_{\Lambda}$. This is in the spirit of original set-up of RNC, where one uses the geodesics of the flat tangent space - the straight lines - to model geodesics in $\mathcal M$. However, as should be clear from our derivations and expressions, one could instead simply use the $\widetilde{\Delta}$ corresponding to $\mathcal{M}$ itself without having to refer to a maximally symmetric tangent space itself - one then simply sets $\widetilde{R}_{\textsf{abcd}} \to {R}_{\textsf{abcd}}$ etc. in all the results. This, incidentally, would yield a local metric which depends on the Ricci tensor of $\mathcal M$ along with the Riemann tensor, and hence, if field equations are imposed, would carry some information about the stress-tensor that is generically not present in the conventional metric in RNC (since stress tensor does not uniquely fix the Riemann tensor). It is at present unclear to us whether this would be a better interpretation. From a purely geometrical point of view, our motivation seems to be extremely close to the one behind Cartan geometry, where the basic idea is to use a maximally symmetric tangent space, and the identification of objects at two different points is then made by rolling this space on the base manifold without slipping. In this context, it is worth pointing out that, in presence of torsion, which is an additional geometrical object present in Cartan's formulation, the auto-parallels will in general be distinct from curves of extremal length. Since all our results use covariant Taylor expansions of derivatives of the world function, torsion will explicitly appear in the expansions (see \cite{cartan-geometry}), and thereby, in the final metric. It will indeed be very interesting to derive these torsion dependent terms in the metric.
This is an elegant generalisation of conventional Riemannian geometry, and it would be worth exploring Cartan geometry\cite{dkw-macdowell,dkw-symmetric-space} using our formalism. It is also worth pointing out that there has been work along similar lines on normal coordinates in the context of Finsler geometry \cite{chris-pf-the_tangent,man-hoh-extensions}. In this context, let us point out an alternate, illuminating way in which our metric, Eq.~(\ref{eq:metric-full}), can be re-expressed after some straightforward manipulations, assuming a maximally symmetric tangent space as in Cartan formulation:
\begin{widetext}
\begin{equation}
	\begin{aligned}
		g_{\textsf{ab}} &= \overset{0}{g}_{\textsf{ab}} + -\frac{1}{3}\left( R_{\textsf a \textsf c \textsf b \textsf d}-\widetilde{R}_{\textsf a \textsf c \textsf b \textsf d} \right)\hat{x}^{\textsf c}\hat{x}^{\textsf d}-\frac{1}{6}\nabla_{\textsf e}R_{\textsf a \textsf c \textsf b \textsf d}\hat{x}^{\textsf c}\hat{x}^{\textsf d}\hat{x}^{\textsf e}+\left\lbrace-\frac{1}{20}\nabla_{\textsf e}\nabla_{\textsf f}R_{\textsf a \textsf c \textsf b \textsf d} \right.\\
		& \quad \;\left. -\frac{2}{9}\widetilde{R}^{\textsf m}_{\phantom{\textsf m}\textsf e \textsf b \textsf f}\left( R_{\textsf a \textsf c \textsf m \textsf d}-\widetilde{R}_{\textsf a \textsf c \textsf m \textsf d} \right)+\frac{2}{45}\left( R_{\textsf a \textsf c \textsf m \textsf d}R^{\textsf m}_{\phantom{m}\textsf e \textsf b \textsf f}-\widetilde{R}_{\textsf a \textsf c \textsf m \textsf d}\widetilde{R}^{\textsf m}_{\phantom{m}\textsf e \textsf b \textsf f} \right) \right\rbrace\hat{x}^{\textsf c}\hat{x}^{\textsf d}\hat{x}^{\textsf e}\hat{x}^{\textsf f}+\mathcal{O}(\textsf{x}^{5})
	\end{aligned}
\end{equation}
\end{widetext}
where 
$
\overset{0}{g}_{\textsf{ab}} = \eta_{\textsf{ab}} + 
		{\widetilde{\Lambda} }\l({1-\widetilde{\Lambda} \eta_{\textsf{e}\textsf{f}}\widehat{x}^{\textsf{e}}\widehat{x}^{\textsf{f}}}\r)^{-1} \eta_{\textsf{a}\textsf{c}}\eta_{\textsf{b} \textsf{d}}\widehat{x}^{\textsf{c}}\widehat{x}^{\textsf{d}} 
$
is the maximally symmetric metric in embedding coordinates; see the discussion in Sec. (\ref{sec:max-symm-sp}), and $\widetilde{R}_{\textsf{abcd}}$ its Riemann tensor.
This form of the metric makes it intuitively very clear that the spacetime geometry described by our metric uses a maximally symmetric tangent space for its local approximation, as in Cartan formulation.

Once the coordinate system based on geodesics of a maximally symmetry tangent space has been appropriately defined, computing the metric is straightforward, albeit lengthy. We have presented here the results of such a computation, highlighting some of the key steps in the derivation along the way. Once the metric in its final form is displayed, one can proceed to analyse physical processes, both classical and quantum, in its background. Since many important physical observables are observer dependent, they will also depend on the curvature of the tangent space. And indeed, the examples we have given already provide illuminating insights. We hope further investigations will shed more light on aspects of local spacetime geometry as characterised in this work.

Several physical effects can be analysed in the background metric that we have derived in this paper, and as future outlook, we list below a few that should be of immediate interest: 
\\
\\
(i) {\it Coupled curvature terms}: The {\it quartic} terms in the metric contain coupled terms involving {\it product} of curvatures of the background spacetime and the tangent space, and it would be interesting to study what new kind of effects such coupled terms can lead to, since they vanish when either the background or the tangent space is flat.
\\
\\
(ii) {\it Implications for quantum dynamics}: It is well known that the van Vleck determinant appears as the pre-factor in the expression for propagation kernel of point particle in the WKB approximation (see, for example, \cite{mcg-chaos_in}). Given this fact, and the manner in which we have defined our local coordinates, the metric we have obtained seems better suited as a background for analysing quantum dynamics. As is well known, in quantum field theory, the choice of coordinates, being tied to the choice of an observer, is crucial since the vacuum state of the theory depends on this choice. Unruh effect is a famous example of this; while one can analyse this effect completely in the Minkowski coordinates, the use of Rindler coordinates not only facilitates computations, but also brings out with much better clarity the role of vacuum fluctuations through the structure of the two-point function expressed in Rindler coordinates. In a similar vein, it will be interesting to ask what kind of vacuum is associated with the coordinates we have defined here. It must be different from the usual Minkowski vacuum, since the ``rest" frames, as we have shown, are accelerated. We hope to present a more complete discussion of these aspects in future work.
\\
\\
(iii) {$\Lambda$ \it as a fundamental constant}: We have already alluded to the idea that our method provides a natural way to weave-in the cosmological constant into the very fabric of spacetime, giving it the status of a fundamental constant\cite{lambda-vacuum-energy}. As mentioned in para 2 above, this is very close in spirit to Cartan geometry, specifically as applied to the MacDowell-Mansouri formulation of general relativity\cite{dkw-macdowell}. It will be worth investigating if/how the condition of {\it rolling without slipping} can be understood in terms of the set-up we have described here. 

\section*{Acknowledgements}
The authors would like to thank Dr Steffen Gielen for a useful correspondence, and the anonymous referee for helpful suggestions that improved clarity of presentation. 
HK would like to thank IIT, Madras and Ministry of Human Resources and Development (MHRD), India for financial support.


\widetext
\appendix

\section{Geometry of equi-geodesic surfaces} \label{app:equi-geodesic} 
In this appendix, we give the induced metric, extrinsic curvature, and intrinsic Ricci scalar of the equi-geodesic surfaces, which by definition comprises of the set of points $p$ at a constant geodesic distance from a given point $p_0$. Such surfaces turn out to be of key significance in characterising the small scale structure of spacetime, and their geometry has been discussed in \cite{equi-geod}. However, the expressions we give below are new, and evaluated in Riemann normal coordinates, which has the following advantage: In covariant Taylor expansions of bi-tensors with both indices at $p$, the coefficients are also evaluated at $p$, and hence, care must be taken while differentiating such series expansions. However, when expressed in RNC, as we do below, the coefficients are all evaluated at $p_0$, and hence the series expansions below are more convenient to use.

Let $x^{\textsf a}$ denote the standard RNC and $u^{\textsf a}(p_{0})$ the normalised tangent vector at the base point; for brevity, we will focus on the case $\eta_{\textsf ab} u^{\textsf a} u^{\textsf b} = -1$. Since $s^{2}=- \eta_{\textsf a \textsf b}x^{\textsf a}x^{\textsf b}$, and $u^{\textsf a}$ can be parametrised in terms of a boost $\chi$ and direction cosines $\theta^A$ ($A=3 \ldots D$) as $u^{\textsf a}\left( \Theta^{\mu} \right)\equiv\left( \cosh \chi, \sinh\chi \, \theta^A \right)$, with $\Theta^\mu \equiv (\chi, \theta^A)$. We therefore change coordinates from $x^{\textsf a} \to (s, \Theta^\mu)$, with $dx^{\textsf a}=u^{\textsf a}\left( \Theta^{\mu} \right)ds+s\Lambda^{\textsf a}_{\mu}d\Theta^{\mu}$, where $\Lambda^{\textsf a}_{\mu}\equiv {\partial u^{\textsf a}}/{\partial\Theta^{\mu}}$. It is now straightforward to substitute this into the line element corresponding to RNC, put $s=$ constant, and thereby read-off the induced metric on the equi-geodesic surface. The final form of the induced metric turns out to be
\begin{eqnarray}
	h_{\mu\nu}=\bar{h}_{\mu\nu}-\frac{1}{3} s^{4} \mathcal{E}_{\mu\nu}-\frac{1}{6} s^{5} \Omega_{\mu\nu}+O(s^{6})\;
\end{eqnarray}
where $\bar{h}_{\mu\nu}=s^{2}\eta_{ab}\Lambda^{a}_{\mu}\Lambda^{b}_{\nu}$ is the induced metric on equi-geodesic surface of Minkowski space, 
$\mathcal{E}_{\mu\nu}=R_{\textsf a\textsf c\textsf b\textsf d}\Lambda^{\textsf a}_{\mu}\Lambda^{\textsf b}_{\nu}u^{c}u^{d}$, and $\Omega_{\mu\nu}=R_{\textsf a\textsf c\textsf b\textsf d;\textsf e}\Lambda^{\textsf a}_{\mu}\Lambda^{\textsf b}_{\nu}u^{c}u^{d}u^{e}$.

Further, it is easy to show that the metric in RNC, when expressed in $(s, \Theta^\mu)$ coordinates, yields a metric in the ADM form with $N=1, N^\mu=0$ \cite{book-MTW}. The extrinsic curvature for the equi-geodesic is therefore $K_{\mu\nu}={\partial h_{\mu\nu}}/{\partial s}$, and using the above expansion for $h_{\mu\nu}$, yields
\begin{eqnarray}
	K_{\mu\nu} &=& \frac{1}{s}\bar{h}_{\mu\nu}-\frac{2}{3} s^{3} \mathcal{E}_{\mu\nu}-\frac{5}{12}s^{4}\Omega_{\mu\nu}+O(s^{5})\;
	\\
	K^{\mu}_{\phantom{\mu}\nu} &=& \frac{1}{s}\delta^{\mu}_{\nu}-\frac{1}{3}s\mathcal{E}^{\mu}_{\phantom{\mu}\nu}-\frac{1}{4}s^{2}\Omega^{\mu}_{\phantom{\mu}\nu}+O(s^{3})
\label{extr-curv-1}
\end{eqnarray}
Note that the expansions above are slightly different from the ones in \cite{equi-geod}, precisely because the coefficients in \cite{equi-geod} are evaluated at $p$, while here we have all the coefficients evaluated at $p_0$. This distinction is subtle and important, particularly when one is dealing with expansions of tensors, and the expansions in RNC might be more convenient to use. For the sake of completeness, we also quote the intrinsic Ricci scalar of the equi-geodesic surfaces:
\begin{eqnarray}
\mathcal R_{\Sigma} &=& -\frac{D_1 D_2}{s^2} + R + \frac{2(D+1)}{3} R_{ab}u^a u^b + O(s)
\label{eq:equi-geod-ricci}
\end{eqnarray}
with all coefficients on the right evaluated at $p_0$.


\section{Derivation of Eq.~(\ref{eq:vvd-jacobian})} \label{app1}

We essentially need to compute the determinant of the matrix 
\begin{eqnarray}
{\mathcal D}_{a'b} = s_0 {\mathcal C}_{a'b} + \varepsilon u_{0a'} u_{b}
\end{eqnarray}
where ${\mathcal C}_{a'b}$ is non-invertible, since it has a zero eigenvalue. To do this, we can use the {\it Matrix determinant lemma}, but we sketch a derivation which naturally yields a relation between scalarised determinants, rather than a relation between tensor densities. 

The computation goes as follows: 

Let $e^{i'}_{\textsf a'}$ and $e^{i}_{\textsf a}$ be tetrads at $p_0$ and $p$, such that $e^{i}_{(0)}=u^i$, $u_{0} e^{i}_{(\mu)} = 0$, and $g_{ab} e^{a}_{(\mu)} e^{b}_{(\nu)} = {\textsf h}_{\bm{\mu \nu}}$, with similar conditions imposed on tetrads at $p_0$. Therefore, the metric in this frame looks like
\begin{eqnarray}
g_{\textsf{ab}} = \left[ \begin{array}{c|c}
\varepsilon & 0
\\
\hline
\\
0 & {\textsf h}_{\bm{\mu \nu}} 
\end{array} 
\right] \;
\end{eqnarray}
from which we immediately obtain ${\rm det}[g_{ij}] {\rm det}[e^{i}_{\textsf a}]^2 = \varepsilon {\rm det}[{\textsf h}_{\bm{\mu \nu}}]$. That is, 
$$
{\rm det}[e^{i}_{\textsf a}] = \sqrt{ \frac{ \varepsilon {\rm det}[{\textsf h}_{\bm{\mu \nu}}] }{{\rm det}[g_{ij}] } }
$$
We therefore have 
\begin{eqnarray}
 {\rm det} [{\mathcal D}_{\textsf a' \textsf b}]
 &=& 
 \sqrt{ \frac{ |{\rm det}[{\textsf h'}_{\bm{\mu \nu}}]| }{-{\rm det}[g'_{ij}] } } \sqrt{ \frac{ |{\rm det}[{\textsf h}_{\bm{\mu \nu}}]| }{- {\rm det}[g_{ij}] } } {\rm det} [{\mathcal D}_{i'j}] 
\nn \\
 &=& s_0^{D-1}{\rm det} [{\mathcal C}_{\bm{\mu'\nu}}]
 \nn
\end{eqnarray}
which, when re-arranged, gives a relation between scalar quantities
\begin{eqnarray}
\underbrace{
\frac{{\rm det} [{\mathcal D}_{a'b}]}{ \sqrt{|g'(p_0)|} \sqrt{|g(p)|} } 
}_{\Delta(p_0, p) }
= \frac{s_0^{D-1} {\rm det} [{\mathcal C}_{\mu' \nu}]}{ \sqrt{|h'(p_0)|} \sqrt{|h(p)|} }
\label{eq:vvd-jacobian}
\end{eqnarray}
\\
where the RHS has now been expressed in arbitrary coordinates on the $(D-1)$ surfaces orthogonal to $u^i$, $u^{i'}$. As indicated, the LHS defines the so called van Vleck bi-scalar $\Delta(p_0, p)$.
\\
\section{Derivation of metric in Eq.~(\ref{eq:metric-full})} \label{app2}
We give below the series expansion for various bi-tensors used in the text (see Ref \cite{smc-vacc}):
\begin{equation}
	\begin{aligned}
		& \sigma^{a'}_{b} =-g^{b'}_{\phantom{b'}b}\left[\delta^{a'}_{b'}+\frac{1}{6}R^{a'}_{\phantom{a'} c' b' d'}\sigma^{c'}\sigma^{d'}-\frac{1}{12}R^{a'}_{\phantom{a'}c' b' d';e'}\sigma^{c'}\sigma^{d'}\sigma^{e'}+\left(\frac{1}{40}R^{a'}_{\phantom{a'}c' b' d';e'f'}
		-\frac{7}{360}R^{a'}_{c'l'd'}R^{l'}_{e'b'f'}\right)\sigma^{c'}\sigma^{d'}\sigma^{e'}\sigma^{f'} +O(\textsf{x}^5)\right] \\
		& \Delta =1+\frac{1}{6}R_{a' b'}\sigma^{a'}\sigma^{b'}-\frac{1}{12}R_{a'b';c'}\sigma^{a'}\sigma^{b'}\sigma^{c'}+\left(\frac{1}{40}R_{a'b';c'd'}+\frac{1}{180}R^{l'}_{\phantom{l'}a' m' b'}R^{m'}_{\phantom{m'}c'l'd'} 
		+\frac{1}{72}R_{a'b}R_{c'd'}\right)\sigma^{a'}\sigma^{b'}\sigma^{c'}\sigma^{d'} +O(\textsf{x}^5) \\
		& \Delta^{p} =1+\frac{p}{6}R_{a' b'}\sigma^{a'}\sigma^{b'}-\frac{p}{12}R_{a'b';c'}\sigma^{a'}\sigma^{b'}\sigma^{c'}+\left(\frac{p}{40}R_{a'b';c'd'}+\frac{p}{180}R^{l'}_{\phantom{l'}a' m' b'}R^{m'}_{\phantom{m'}c'l'd'} 
		+\frac{p^{2}}{72}R_{a'b}R_{c'd'}\right)\sigma^{a'}\sigma^{b'}\sigma^{c'}\sigma^{d'} +O(\textsf{x}^5) \\	
		& \ln \Delta =\frac{1}{6}R_{a' b'}\sigma^{a'}\sigma^{b'}-\frac{1}{12}R_{a'b';c'}\sigma^{a'}\sigma^{b'}\sigma^{c'}+\left(\frac{1}{40}R_{a'b';c'd'} 
		+\frac{1}{180}R^{l'}_{\phantom{l'}a' m' b'}R^{m'}_{\phantom{m'}c'l'd'}\right)\sigma^{a'}\sigma^{b'}\sigma^{c'}\sigma^{d'}+O(\textsf{x}^5)
	\end{aligned}
\label{series}
\end{equation}
where $g^{b'}_{\phantom{b'}b}\equiv e^{b'}_{a}e^{a}_{b}$ is the parallel propagator and $p$ is an integer.

Using the transformation law given in Eq.~(\ref{coord-trans}), the Riemann tensor transforms as, $ R^{a'}_{\phantom{a'}c'b'd'} =R^{\textsf{a}}_{\phantom{\textsf a}\textsf{c}\textsf{b}\textsf{d}}e^{a'}_{\textsf{a}}e^{\textsf{c}}_{c'}e^{\textsf{b}}_{b'}e^{\textsf{d}}_{d'}$. Substitute this transformation into the expansion and using it to the result of Eq.~(\ref{coord-var}), the variation in the local coordinates become,
\begin{equation}
	\begin{aligned}
		\DM \widehat{x}^{\textsf{a}} &=\left\lbrace \left[1-\frac{\widetilde{\Delta}^{2/D_{1}}}{6D_{1}}\widetilde{R}_{\textsf{c}\textsf{d}}\widehat{x}^{\textsf{c}}\widehat{x}^{\textsf{d}}+\frac{\widetilde{\Delta}^{3/D_{1}}}{12D_{1}}\widetilde{R}_{\textsf{c}\textsf{d};\textsf{e}}\widehat{x}^{\textsf{c}}\widehat{x}^{\textsf{d}}\widehat{x}^{\textsf{e}}-\frac{\widetilde{\Delta}^{4/D_{1}}}{D_{1}}\left(\frac{1}{40}\widetilde{R}_{\textsf{c}\textsf{d};\textsf{e}\textsf{f}}+\frac{1}{180}\widetilde{R}^{\textsf{l}}_{\phantom{\textsf l}\textsf{c}\textsf{m}\textsf{d}}\widetilde{R}^{\textsf{m}}_{\phantom{\textsf m}\textsf{e}\textsf{l}\textsf{f}}-\frac{1}{72D_{1}}\widetilde{R}_{\textsf{c}\textsf{d}}\widetilde{R}_{\textsf{e}\textsf{f}}\right)\widehat{x}^{\textsf{c}}\widehat{x}^{\textsf{d}}\widehat{x}^{\textsf{e}}\widehat{x}^{\textsf{f}}\right]\right. \\
	 	&\quad \quad \left. \left[\delta^{\textsf{a}}_{\textsf{r}}+\frac{\widetilde{\Delta}^{2/D_{1}}}{6}R^{\textsf{a}}_{\phantom{\textsf a}\textsf{c}\textsf{r}\textsf{d}}\widehat{x}^{\textsf{c}}\widehat{x}^{\textsf{d}}-\frac{\widetilde{\Delta}^{3/D_{1}}}{12}R^{\textsf{a}}_{\phantom{\textsf a}\textsf{c}\textsf{r}\textsf{d};\textsf{e}}\widehat{x}^{\textsf{c}}\widehat{x}^{\textsf{d}}\widehat{x}^{\textsf{e}}+\left(\frac{1}{40}R^{\textsf{a}}_{\phantom{\textsf a}\textsf{c}\textsf{r}\textsf{d};\textsf{e}\textsf{f}}+\frac{7}{360}R^{\textsf{a}}_{\phantom{\textsf a}\textsf{c}\textsf{i}\textsf{d}}R^{\textsf{i}}_{\phantom{\textsf i}\textsf{e}\textsf{r}\textsf{f}}\right)\widehat{x}^{\textsf{c}}\widehat{x}^{\textsf{d}}\widehat{x}^{\textsf{e}}\widehat{x}^{\textsf{f}}\widetilde{\Delta}^{4/D_{1}}\right] e^{\textsf{r}}_{b}  \right.\\
	 	&\quad \quad \left. -\frac{\widehat{x}^{\textsf{a}}}{D_{1}}\partial_{b}\ln\left(\widetilde{\Delta}\right) +O(\textsf{x}^5)\right\rbrace \DM x^{b}
	\end{aligned}
\end{equation}
The last term which contains the derivative of $\ln\left(\widetilde{\Delta}\right)$ needs to be expanded. The expansion coefficients themselves depend on $\widetilde{\Delta}$ since our definition 
of coordinates involve $\widetilde{\Delta}$, and hence one must deal with the Taylor expansion recursively to obtain the result at required order. This is perhaps the only part in the derivation which requires careful handling.

To the fourth order the variation becomes,
\begin{equation}
	\begin{aligned}
		\DM \widehat{x}^{\textsf{a}} &=\left\lbrace \delta^{\textsf{a}}_{\textsf{r}}+\left(\frac{1}{6}R^{\textsf{a}}_{\phantom{\textsf a}\textsf{c}\textsf{r}\textsf{d}}-\frac{\delta^{\textsf{a}}_{\textsf{d}}}{3D_{1}}\widetilde{R}_{\textsf{r}\textsf{c}}-\frac{\delta^{\textsf{a}}_{\textsf{r}}}{6D_{1}}\widetilde{R}_{\textsf{c}\textsf{d}}\right)\widehat{x}^{\textsf{c}}\widehat{x}^{\textsf{d}}+\left(\frac{1}{12}R^{\textsf{a}}_{\phantom{\textsf a}\textsf{c}\textsf{r}\textsf{d};\textsf{e}}+\frac{\delta^{\textsf{a}}_{\textsf{r}}}{12D_{1}}\widetilde{R}_{\textsf{c}\textsf{d};\textsf{e}}-\frac{\delta^{\textsf{a}}_{\textsf{e}}}{4D_{1}}\widetilde{R}_{\textsf{c}\textsf{d};\textsf{r}}\right)\widehat{x}^{\textsf{c}}\widehat{x}^{\textsf{d}}\widehat{x}^{\textsf{e}}\right.\\ 
		& \quad \;\; \left. +\left(\frac{1}{40}R^{\textsf{a}}_{\phantom{\textsf a}\textsf{c}\textsf{r}\textsf{d};\textsf{e}\textsf{f}}+\frac{7}{360}R^{\textsf{a}}_{\phantom{\textsf a}\textsf{c}\textsf{l}\textsf{d}}R^{\textsf{l}}_{\phantom{\textsf l}\textsf{e}\textsf{r}\textsf{f}}+\frac{1}{36D_{1}}R^{\textsf{a}}_{\phantom{\textsf a}\textsf{c}\textsf{r}\textsf{d}}\widetilde{R}_{\textsf{e}\textsf{f}}-\frac{\delta^{\textsf{a}}_{\textsf{f}}}{6D_{1}}R^{\textsf{l}}_{\phantom{\textsf l}\textsf{c}\textsf{r}\textsf{d}}\widetilde{R}_{\textsf{l}\textsf{e}}-\frac{\delta^{\textsf{a}}_{\textsf{r}}}{40D_{1}}\widetilde{R}_{\textsf{c}\textsf{d};\textsf{e}\textsf{f}}-\frac{\delta^{\textsf{a}}_{\textsf{f}}}{10D_{1}}\widetilde{R}_{\textsf{c}\textsf{d};\textsf{e}\textsf{r}}\right.\right.\\ 
		& \quad \;\; \left.\left.-\frac{\delta^{\textsf{a}}_{\textsf{r}}}{24D_{1}^2}\widetilde{R}_{\textsf{c}\textsf{d}}\widetilde{R}_{\textsf{e}\textsf{f}}-\frac{\delta^{\textsf{a}}_{\textsf{f}}}{18D_{1}^2}\widetilde{R}_{\textsf{c}\textsf{r}}\widetilde{R}_{\textsf{e}\textsf{d}}-\frac{\delta^{\textsf{a}}_{\textsf{r}}}{180D_{1}}\widetilde{R}^{\textsf{l}}_{\phantom{\textsf l}\textsf{c}\textsf{m}\textsf{d}}\widetilde{R}^{\textsf{m}}_{\phantom{\textsf m}\textsf{e}\textsf{l}\textsf{f}}-\frac{\delta^{\textsf{a}}_{\textsf{f}}}{45D_{1}}\widetilde{R}^{\textsf{l}}_{\phantom{\textsf l}\textsf{c}\textsf{m}\textsf{d}}\widetilde{R}^{\textsf{m}}_{\phantom{\textsf m}\textsf{e}\textsf{l}\textsf{r}} \right) \widehat{x}^{\textsf{c}}\widehat{x}^{\textsf{d}}\widehat{x}^{\textsf{e}}\widehat{x}^{\textsf{f}}+O(\textsf{x}^5)\right\rbrace e^{\textsf{r}}_{a}\DM x^{a}
	\end{aligned}
\end{equation}
This expansion is inverted to find the line element in the local coordinates as,
\begin{equation}
	\begin{aligned}
		e^{\textsf{a}}_{a}\DM x^{a} &=\left\lbrace \delta^{\textsf{a}}_{\textsf{r}}+\left(-\frac{1}{6}R^{\textsf{a}}_{\phantom{\textsf a}\textsf{c}\textsf{r}\textsf{d}}+\frac{\delta^{\textsf{a}}_{\textsf{d}}}{3D_{1}}\widetilde{R}_{\textsf{r}\textsf{c}}+\frac{\delta^{\textsf{a}}_{r}}{6D_{1}}\widetilde{R}_{\textsf{c}\textsf{d}}\right)\widehat{x}^{\textsf{c}}\widehat{x}^{\textsf{d}}+\left( -\frac{1}{12}R^{\textsf{a}}_{\phantom{\textsf a}\textsf{c}\textsf{r}\textsf{d};\textsf{e}}-\frac{\delta^{\textsf{a}}_{\textsf{r}}}{12D_{1}}\widetilde{R}_{\textsf{c}\textsf{d};\textsf{e}}+\frac{\delta^{\textsf{a}}_{\textsf{e}}}{4D_{1}}\widetilde{R}_{\textsf{c}\textsf{d};\textsf{r}} \right)\widehat{x}^{\textsf{c}}\widehat{x}^{\textsf{d}}\widehat{x}^{\textsf{e}} \right.\\  
		& \quad\;\; \left. +\left( -\frac{1}{40}R^{\textsf{a}}_{\phantom{\textsf a}\textsf{c}\textsf{r}\textsf{d};\textsf{e}\textsf{f}}-\frac{7}{360}R^{\textsf{a}}_{\phantom{\textsf a}\textsf{c}\textsf{l}\textsf{d}}R^{\textsf{l}}_{\phantom{\textsf l}\textsf{e}\textsf{r}\textsf{f}}-\frac{1}{36D_{1}}R^{\textsf{a}}_{\phantom{\textsf a}\textsf{c}\textsf{r}\textsf{d}}\widetilde{R}_{\textsf{e}\textsf{f}}+\frac{\delta^{\textsf{a}}_{\textsf{f}}}{6D_{1}}R^{\textsf{l}}_{\phantom{\textsf l}\textsf{\textsf{c}}\textsf{r}\textsf{\textsf{d}}}\widetilde{R}_{\textsf{l}\textsf{e}}+\frac{\delta^{\textsf{a}}_{\textsf{r}}}{40D_{1}}\widetilde{R}_{\textsf{c}\textsf{\textsf{d}};\textsf{\textsf{e}}\textsf{\textsf{f}}}+\frac{\delta^{\textsf{a}}_{\textsf{f}}}{10D_{1}}\widetilde{R}_{\textsf{c}\textsf{\textsf{d}};\textsf{\textsf{e}}\textsf{r}} \right. \right.\\ 
		& \quad\;\; \left.\left. +\frac{\delta^{\textsf{a}}_{\textsf{r}}}{24D_{1}^2}\widetilde{R}_{\textsf{c}\textsf{\textsf{d}}}\widetilde{R}_{\textsf{\textsf{e}}\textsf{\textsf{f}}}+\frac{\delta^{\textsf{a}}_{\textsf{f}}}{18D_{1}^2}\widetilde{R}_{\textsf{c}\textsf{r}}\widetilde{R}_{\textsf{\textsf{e}}\textsf{\textsf{d}}}+\frac{\delta^{\textsf{a}}_{\textsf{r}}}{180D_{1}}\widetilde{R}^{\textsf{l}}_{\phantom{\textsf l}\textsf{c}\textsf{m}\textsf{\textsf{d}}}\widetilde{R}^{\textsf{m}}_{\phantom{\textsf m}\textsf{\textsf{e}}\textsf{l}\textsf{\textsf{f}}}+\frac{\delta^{\textsf{a}}_{\textsf{f}}}{45D_{1}}\widetilde{R}^{\textsf{l}}_{\phantom{\textsf l}\textsf{c}\textsf{m}\textsf{\textsf{d}}}\widetilde{R}^{\textsf{m}}_{\phantom{\textsf m}\textsf{\textsf{e}}\textsf{l}\textsf{r}} +\frac{1}{36}R^{\textsf{a}}_{\phantom{\textsf a}\textsf{\textsf{c}}\textsf{l}\textsf{\textsf{d}}}R^{\textsf{l}}_{\phantom{\textsf l}\textsf{\textsf{e}}\textsf{r}\textsf{\textsf{f}}} \right. \right.\\ 
		& \quad\;\; \left.\left. -\frac{1}{18D_{1}}R^{\textsf{a}}_{\phantom{\textsf a}\textsf{\textsf{c}}\textsf{\textsf{f}}\textsf{\textsf{d}}}\widetilde{R}_{\textsf{\textsf{e}}\textsf{r}} -\frac{1}{36D_{1}}R^{\textsf{a}}_{\phantom{\textsf a}\textsf{\textsf{c}}\textsf{r}\textsf{\textsf{d}}}\widetilde{R}_{\textsf{\textsf{e}}\textsf{\textsf{f}}}-\frac{\delta^{\textsf{a}}_{\textsf{d}}}{18D_{1}}R^{\textsf{l}}_{\phantom{\textsf l}\textsf{\textsf{e}}\textsf{r}\textsf{\textsf{f}}}\widetilde{R}_{\textsf{\textsf{c}}\textsf{l}}+\frac{\delta^{\textsf{a}}_{\textsf{d}}}{9D_{1}^2}\widetilde{R}_{\textsf{\textsf{c}}\textsf{\textsf{f}}}\widetilde{R}_{\textsf{\textsf{e}}\textsf{r}}+\frac{\delta^{\textsf{a}}_{\textsf{d}}}{18D_{1}^2}\widetilde{R}_{\textsf{\textsf{c}}\textsf{r}}\widetilde{R}_{\textsf{\textsf{e}}\textsf{\textsf{f}}}-\frac{1}{36D_{1}}R^{\textsf{a}}_{\phantom{\textsf a}\textsf{\textsf{e}}\textsf{r}\textsf{\textsf{f}}}\widetilde{R}_{\textsf{\textsf{c}}\textsf{\textsf{d}}} \right. \right.\\ 
		& \quad\;\; \left.\left.+\frac{\delta^{\textsf{a}}_{\textsf{f}}}{18D_{1}^2}\widetilde{R}_{\textsf{\textsf{c}}\textsf{\textsf{d}}}\widetilde{R}_{\textsf{\textsf{e}}\textsf{r}}+\frac{\delta^{\textsf{a}}_{\textsf{r}}}{36D_{1}^2}\widetilde{R}_{\textsf{\textsf{c}}\textsf{\textsf{d}}}\widetilde{R}_{\textsf{\textsf{e}}\textsf{\textsf{f}}}\right)\widehat{x}^{\textsf{c}}\widehat{x}^{\textsf{d}}\widehat{x}^{\textsf{e}}\widehat{x}^{\textsf{f}}+O(\textsf{x}^5) \right\rbrace \DM \widehat{x}^{\textsf a}\; .
	\end{aligned}
\end{equation}
The line element is evaluated using the definition, $\DM s^{2}=\eta_{\textsf{a}\textsf{b}}e^{\textsf{a}}_{a}e^{\textsf{b}}_{b}\DM x^{a}\DM x^{b}$ and we have,
\begin{equation}
	\begin{aligned}
		\DM \sigma^2 &= \left\lbrace \eta_{\textsf{r}\textsf{s}}+\left( -\frac{\eta_{\textsf{a}\textsf{s}}}{3}R^{\textsf{a}}_{\phantom{\textsf a}\textsf{c}\textsf{r}\textsf{d}}+\frac{\eta_{\textsf{r}\textsf{s}}}{3D_{1}}\widetilde{R}_{\textsf{c}\textsf{d}}+\frac{\eta_{\textsf{d}\textsf{s}}}{3D_{1}}\widetilde{R}_{\textsf{c}\textsf{r}}+\frac{\eta_{\textsf{r}\textsf{d}}}{3D_{1}}\widetilde{R}_{\textsf{c}\textsf{s}} \right)\widehat{x}^{\textsf{c}}\widehat{x}^{\textsf{d}}+\left( -\frac{\eta_{\textsf{a}\textsf{s}}}{6}R^{\textsf{a}}_{\phantom{\textsf a}\textsf{c}\textsf{r}\textsf{d};\textsf{e}}-\frac{\eta_{\textsf{r}\textsf{s}}}{6D_{1}}\widetilde{R}_{\textsf{c}\textsf{d};\textsf{e}}+\frac{\eta_{\textsf{e}\textsf{s}}}{4D_{1}}\widetilde{R}_{\textsf{c}\textsf{d};\textsf{r}} \right. \right.\\ 
		& \quad\;\; \left. \left. +\frac{\eta_{\textsf{r}\textsf{e}}}{4D_{1}}\widetilde{R}_{\textsf{c}\textsf{d};\textsf{s}}\right)\widehat{x}^{\textsf{c}}\widehat{x}^{\textsf{d}}\widehat{x}^{\textsf{e}}+\left( -\frac{\eta_{\textsf{a}\textsf{s}}}{20}R^{\textsf{a}}_{\phantom{\textsf a}\textsf{c}\textsf{r}\textsf{d};\textsf{e}\textsf{f}}+\frac{2\eta_{\textsf{a}\textsf{s}}}{45}R^{\textsf{a}}_{\phantom{\textsf a}\textsf{c}\textsf{l}\textsf{d}}R^{\textsf{l}}_{\phantom{\textsf l}\textsf{e}\textsf{r}\textsf{f}}-\frac{2\eta_{\textsf{a}\textsf{s}}}{9D_{1}}R^{\textsf{a}}_{\phantom{\textsf a}\textsf{c}\textsf{r}\textsf{d}}\widetilde{R}_{\textsf{e}\textsf{f}}+\frac{\eta_{\textsf{f}\textsf{s}}}{9D_{1}}R^{\textsf{l}}_{\phantom{\textsf l}\textsf{c}\textsf{r}\textsf{d}}\widetilde{R}_{\textsf{l}\textsf{e}}+\frac{\eta_{\textsf{r}\textsf{f}}}{9D_{1}}R^{\textsf{l}}_{\phantom{\textsf l}\textsf{c}\textsf{s}\textsf{d}}\widetilde{R}_{\textsf{l}\textsf{e}} \right. \right.\\ 
		& \quad\;\; \left. \left. +\frac{\eta_{\textsf{r}\textsf{s}}}{20D_{1}}\widetilde{R}_{\textsf{c}\textsf{d};\textsf{e}\textsf{f}}+\frac{\eta_{\textsf{f}\textsf{s}}}{10D_{1}}\widetilde{R}_{\textsf{c}\textsf{d};\textsf{e}\textsf{r}}+\frac{\eta_{\textsf{r}\textsf{f}}}{10D_{1}}\widetilde{R}_{\textsf{c}\textsf{d};\textsf{e}\textsf{s}}+\frac{\eta_{\textsf{r}\textsf{s}}}{6D_{1}^2}\widetilde{R}_{\textsf{c}\textsf{d}}\widetilde{R}_{\textsf{e}\textsf{f}}+\frac{\eta_{\textsf{f}\textsf{s}}}{3D_{1}^2}\widetilde{R}_{\textsf{c}\textsf{d}}\widetilde{R}_{\textsf{e}\textsf{r}}+\frac{\eta_{\textsf{r}\textsf{f}}}{3D_{1}^2}\widetilde{R}_{\textsf{c}\textsf{d}}\widetilde{R}_{\textsf{e}\textsf{s}}+\frac{\eta_{\textsf{d}\textsf{f}}}{9D_{1}^2}\widetilde{R}_{\textsf{c}\textsf{r}}\widetilde{R}_{\textsf{e}\textsf{s}} \right. \right.\\ 
		& \quad\;\; \left. \left. +\frac{\eta_{\textsf{r}\textsf{s}}}{90D_{1}}\widetilde{R}^{\textsf{l}}_{\phantom{\textsf l}\textsf{c}\textsf{m}\textsf{d}}\widetilde{R}^{\textsf{m}}_{\phantom{\textsf m}\textsf{e}\textsf{l}\textsf{f}}+\frac{\eta_{\textsf{f}\textsf{s}}}{45D_{1}}\widetilde{R}^{\textsf{l}}_{\phantom{\textsf l}\textsf{c}\textsf{m}\textsf{d}}\widetilde{R}^{\textsf{m}}_{\phantom{\textsf m}\textsf{e}\textsf{l}\textsf{r}}+\frac{\eta_{\textsf{r}\textsf{f}}}{45D_{1}}\widetilde{R}^{\textsf{l}}_{\phantom{\textsf l}\textsf{c}\textsf{m}\textsf{d}}\widetilde{R}^{\textsf{m}}_{\phantom{\textsf m}\textsf{e}\textsf{l}\textsf{s}} \right)\widehat{x}^{\textsf{c}}\widehat{x}^{\textsf{d}}\widehat{x}^{\textsf{e}}\widehat{x}^{\textsf{f}}+O(\textsf{x}^5) \right\rbrace \DM \widehat{x}^{\textsf{r}}\DM \widehat{x}^{\textsf{s}}\;
	\end{aligned}
\end{equation}
Finally we arrive at the metric to quartic order for an arbitrary tangent space as given in Eq.~(\ref{eq:metric-full}).

\section{Inverse metric, determinant and Christoffel symbols} \label{app3}

Let the form of inverse metric be, 
\begin{equation}
\label{eq16_sec3eq16}
	g^{\textsf{a}\textsf{b}}=\eta^{\textsf{a}\textsf{b}}+\text{F}^{\textsf{a}\textsf{b}}_{\phantom{\textsf a\textsf b}\textsf{c}\textsf{d}}\hat{x}^{\textsf{c}}\hat{x}^{\textsf{d}}+\text{G}^{\textsf{a}\textsf{b}}_{\phantom{\textsf a\textsf b}\textsf{c}\textsf{d}\textsf{e}}\hat{x}^{\textsf{c}}\hat{x}^{\textsf{d}}\hat{x}^{\textsf{e}}+\text{H}^{\textsf{a}\textsf{b}}_{\phantom{\textsf a\textsf b}\textsf{c}\textsf{d}\textsf{e}\textsf{f}}\hat{x}^{\textsf{c}}\hat{x}^{\textsf{d}}\hat{x}^{\textsf{e}}\hat{x}^{\textsf{f}}+O(\textsf{x}^{5})\;,
\end{equation}
where the tensor coefficients  $\text{F}^{\textsf{a}\textsf{b}}_{\phantom{\textsf a\textsf b}\textsf{c}\textsf{d}}$, $\text{G}^{\textsf{a}\textsf{b}}_{\phantom{\textsf a\textsf b}\textsf{c}\textsf{d}\textsf{e}}$ and $\text{H}^{\textsf{a}\textsf{b}}_{\phantom{\textsf a\textsf b}\textsf{c}\textsf{d}\textsf{e}\textsf{f}}$ need to be found. These coefficients can be calculated by using the identity $g_{\textsf{a}\textsf{b}}g^{\textsf{c}\textsf{b}}=\delta^{\textsf{c}}_{\textsf{a}}$ and demanding that all the higher order terms in the expansion of this contraction will be zero in every order so that, only the $\eta$ part will contribute to Kronecker delta. The inverse metric is given by,
\begin{equation}
	\begin{aligned}
		g^{\textsf{a}\textsf{b}} &=\eta^{\textsf{a}\textsf{b}}+\frac{1}{3}\left(\eta^{\textsf{a}\textsf{l}} R^{\textsf{b}}_{\phantom{\textsf b}\textsf{c}\textsf{l}\textsf{d}}-\frac{1}{D_{1}}\eta^{\textsf{a}\textsf{b}}\widetilde{R}_{\textsf{c}\textsf{d}}-\frac{1}{D_{1}}\delta^{\textsf{b}}_{\textsf{d}}\eta^{\textsf{a}\textsf{l}}\widetilde{R}_{\textsf{l}\textsf{c}}-\frac{1}{D_{1}}\delta^{\textsf{a}}_{\textsf{d}}\eta^{\textsf{b}\textsf{l}}\widetilde{R}_{\textsf{l}\textsf{c}}\right)\widehat{x}^{\textsf{c}}\widehat{x}^{\textsf{d}}+\left( \frac{1}{6}\eta^{\textsf{a}\textsf{l}}R^{\textsf{b}}_{\phantom{\textsf b}\textsf{c}\textsf{l}\textsf{d};\textsf{e}}+\frac{1}{6D_{1}}\eta^{\textsf{a}\textsf{b}}\widetilde{R}_{\textsf{c}\textsf{d};\textsf{e}} \right. \\
		& \quad \; \left. -\frac{1}{4D_{1}}\delta^{\textsf{b}}_{\textsf{e}}\eta^{\textsf{a}\textsf{l}}\widetilde{R}_{\textsf{c}\textsf{d};\textsf{l}}-\frac{1}{4D_{1}}\delta^{\textsf{a}}_{\textsf{e}}\eta^{\textsf{b}\textsf{l}}\widetilde{R}_{\textsf{c}\textsf{d};\textsf{l}} \right)\widehat{x}^{\textsf{c}}\widehat{x}^{\textsf{d}}\widehat{x}^{\textsf{e}}+\left( \frac{1}{20}\eta^{\textsf{a}\textsf{i}}R^{\textsf{b}}_{\phantom{\textsf b}\textsf{c}\textsf{i}\textsf{d};\textsf{e}\textsf{f}}+\frac{1}{15}\eta^{\textsf{a}\textsf{i}}R^{\textsf{b}}_{\phantom{\textsf b}\textsf{c}\textsf{l}\textsf{d}}R^{\textsf{l}}_{\phantom{\textsf l}\textsf{e}\textsf{i}\textsf{f}}-\frac{2}{9D_{1}}\delta^{\textsf{b}}_{\textsf{f}}\eta^{\textsf{a}\textsf{i}}R^{\textsf{l}}_{\phantom{\textsf l}\textsf{c}\textsf{i}\textsf{d}}\widetilde{R}_{\textsf{l}\textsf{e}} \right.\\
		& \quad \; \left. -\frac{1}{9D_{1}}\delta^{\textsf{a}}_{\textsf{f}}\eta^{\textsf{l}\textsf{m}}R^{\textsf{b}}_{\phantom{\textsf b}\textsf{c}\textsf{l}\textsf{d}}\widetilde{R}_{\textsf{e}\textsf{m}}-\frac{1}{9D_{1}}\delta^{\textsf{a}}_{\textsf{f}}\eta^{\textsf{b}\textsf{m}}R^{\textsf{l}}_{\phantom{\textsf l}\textsf{c}\textsf{m}\textsf{d}}\widetilde{R}_{\textsf{l}\textsf{e}}-\frac{1}{20D_{1}}\eta^{\textsf{a}\textsf{b}}\widetilde{R}_{\textsf{c}\textsf{d};\textsf{e}\textsf{f}}-\frac{1}{10D_{1}}\delta^{\textsf{b}}_{\textsf{f}}\eta^{\textsf{a}\textsf{l}}\widetilde{R}_{\textsf{c}\textsf{d};\textsf{e}\textsf{l}}-\frac{1}{10D_{1}}\delta^{\textsf{a}}_{\textsf{f}}\eta^{\textsf{b}\textsf{l}}\widetilde{R}_{\textsf{c}\textsf{d};\textsf{e}\textsf{l}} \right.\\ 
		& \quad \; \left. -\frac{1}{18D_{1}^2}\eta^{\textsf{a}\textsf{b}}\widetilde{R}_{\textsf{c}\textsf{d}}\widetilde{R}_{\textsf{e}\textsf{f}}-\frac{1}{9D_{1}^2}\delta^{\textsf{b}}_{\textsf{f}}\delta^{\textsf{a}}_{\textsf{c}}\eta^{\textsf{l}\textsf{m}}\widetilde{R}_{\textsf{d}\textsf{m}}\widetilde{R}_{\textsf{e}\textsf{l}}-\frac{1}{90D_{1}}\eta^{\textsf{a}\textsf{b}}\widetilde{R}^{\textsf{l}}_{\phantom{\textsf l}\textsf{c}\textsf{m}\textsf{d}}\widetilde{R}^{\textsf{m}}_{\phantom{\textsf m}\textsf{e}\textsf{l}\textsf{f}}-\frac{1}{45D_{1}}\delta^{\textsf{b}}_{\textsf{f}}\eta^{\textsf{a}\textsf{i}}\widetilde{R}^{\textsf{l}}_{\phantom{\textsf l}\textsf{c}\textsf{m}\textsf{d}}\widetilde{R}^{\textsf{m}}_{\phantom{\textsf m}\textsf{e}\textsf{l}\textsf{i}} \right. \\
		& \quad \; \left. -\frac{1}{45D_{1}}\delta^{\textsf{a}}_{\textsf{f}}\eta^{\textsf{b}\textsf{i}}\widetilde{R}^{\textsf{l}}_{\phantom{\textsf l}\textsf{c}\textsf{m}\textsf{d}}\widetilde{R}^{\textsf{m}}_{\phantom{\textsf m}\textsf{e}\textsf{l}\textsf{i}} \right)\widehat{x}^{\textsf{c}}\widehat{x}^{\textsf{d}}\widehat{x}^{\textsf{e}}\widehat{x}^{\textsf{f}}+O(\textsf{x}^5)\;
  \end{aligned}
\end{equation}

The determinant of the metric can be evaluated by considering the metric in the form $g_{\textsf{a}\textsf{b}}=\eta_{\textsf{a}\textsf{i}}\left( \delta^{\textsf{i}}_{\textsf{b}}+\xi \eta^{\textsf{i}\textsf{j}}A_{\textsf{j}\textsf{b}} \right)$, where $I$ is $\delta^{\textsf{i}}_{\textsf{b}}$ and $A_{\textsf{j}\textsf{b}}$ is the expansion terms. The determinant of the metric $\text{det}\left(g_{\textsf{a}\textsf{b}}\right)=\text{det}\left(\eta_{\textsf{a}\textsf{i}}\right)\text{det}\left( \delta^{\textsf{i}}_{\textsf{b}}+\xi \eta^{\textsf{i}\textsf{j}}A_{\textsf{j}\textsf{b}} \right)$, and determinant of the second term can be related to the trace of $\eta^{\textsf{i}\textsf{j}}A_{\textsf{j}\textsf{b}}$ by
\begin{equation}
\label{eq19_sec3eq19}
	\text{det}(I+\xi A)=1+\xi \text{Tr}A+\frac{\xi^{2}}{2}\left( \left( \text{Tr}A \right)^{2}-\text{Tr}A^{2} \right)+O(\xi^3)
\end{equation}
Trace of the expansion terms of the metric is found by contracting with $\delta^{\textsf{b}}_{\textsf{i}}$. The determinant is then given by
\begin{eqnarray}
	-g & =1 - \frac{1}{3}\left( R_{\textsf{c}\textsf{d}} - \frac{D+2}{D-1}\widetilde{R}_{\textsf{c}\textsf{d}} \right)\widehat{x}^{\textsf{c}}\widehat{x}^{\textsf{d}}-\frac{1}{6}\left( R_{\textsf{c}\textsf{d};\textsf{e}}+\frac{D-3}{D-{1}}\widetilde{R}_{\textsf{c}\textsf{d};\textsf{e}} \right)\widehat{x}^{\textsf{c}}\widehat{x}^{\textsf{d}}\widehat{x}^{\textsf{e}}+O(\textsf{x}^4)\;
\end{eqnarray}
The Christoffel symbols, to leading order, are given by
\begin{eqnarray}
	\Gamma^{\textsf{a}}_{\phantom{\textsf{a}} \textsf{b}\textsf{c}} &=\frac{1}{3}\left\lbrace\left( R^{\textsf{a}}_{\phantom{\textsf a}\textsf{b}\textsf{e}\textsf{c}}+R^{\textsf{a}}_{\phantom{\textsf a}\textsf{c}\textsf{e}\textsf{b}} \right)+\frac{1}{D_{1}}\left( \delta^{\textsf{a}}_{\textsf{b}}\widetilde{R}_{\textsf{c}\textsf{e}}+\delta^{\textsf{a}}_{\textsf{c}}\widetilde{R}_{\textsf{b}\textsf{e}}+\delta^{\textsf{a}}_{\textsf{e}}\widetilde{R}_{\textsf{b}\textsf{c}} \right)\right\rbrace \widehat{x}^{\textsf{e}}+
	O(\textsf{x}^4)\;
\end{eqnarray}


\end{document}